\documentclass[a4paper,prd,twocolumn,nobalancelastpage,superscriptaddress,amsmath,amssymb,oneside]{revtex4-1}
\usepackage[english]{babel}
\usepackage[utf8]{inputenc}
\usepackage[sc]{mathpazo}
\usepackage{graphicx}
\usepackage{rotfloat}
\usepackage{booktabs}
\usepackage{url}
\usepackage{xspace}
\usepackage{xcolor}
\usepackage{todonotes}

\definecolor{darkred}{rgb}{0.5,0,0}
\definecolor{darkblue}{rgb}{0,0,0.5}
\definecolor{firebrick}{rgb}{0.75,0.125,0.125}
\definecolor{darkgreen}{rgb}{0,0.5,0}
\definecolor{darkyellow}{rgb}{0.5,0.5,0}
\definecolor{darkcyan}{rgb}{0,0.5,0.5}
\definecolor{rulecolor}{gray}{0.8}
\usepackage[colorlinks=true,linkcolor=firebrick,citecolor=darkgreen,urlcolor=darkblue]{hyperref}

\renewcommand{\vec}[1]{\mathbf{#1}}
\newcommand{\capt}[2]{\caption{#2}\label{fig:#1}}

\newcommand{\tb}[1]{Table\,\ref{fig:#1}}
\newcommand{\fg}[1]{Fig.\,\ref{fig:#1}}
\newcommand{\eq}[1]{Eq.\,\eqref{eq:#1}}
\newcommand{\dd}{\mathrm{d}}

\newcommand{\pdf}{pdf\xspace}
\newcommand{\pdfs}{pdfs\xspace}
\newcommand{\sest}{{\hat{S}}}
\newcommand{\eest}{{\hat{E}}}
\newcommand{\thest}{\hat{\theta}}
\newcommand{\ms}{{\bar{S}}}
\newcommand{\va}{\vec{a}}
\newcommand{\vp}{\vec{p}}
\newcommand{\vq}{\vec{q}}
\newcommand{\vpest}{\hat{\vp}}
\newcommand{\vaest}{\hat{\va}}
\newcommand{\lnL}{\ln\!\mathcal{L}}
\newcommand{\ecut}{{E_\text{cut}}}
\newcommand{\erfc}{\operatorname{erfc}}

\newcommand{\expect}[1]{\mathrm{E}[#1]}

\newcommand{\normal}{{\mathcal{N}}}
\newcommand{\etal}{{\it et al.}\xspace}

\begin{document}

\title{A likelihood method to cross-calibrate air-shower detectors}

\author{H.P.~Dembinski}
\affiliation{Karlsruhe Institute of Technology}
\affiliation{University of Delaware}

\author{B.~K\'egl}
\affiliation{Laboratoire de l'Acc\'el\'erateur Lin\'eaire}

\author{I.C.~Mari\c{s}}
\affiliation{Universidad de Granada}

\author{M.~Roth}
\affiliation{Karlsruhe Institute of Technology}

\author{D.~Veberi\v{c}}
\affiliation{Laboratoire de l'Acc\'el\'erateur Lin\'eaire}
\affiliation{Karlsruhe Institute of Technology}

\begin{abstract}
We present a detailed statistical treatment of the energy calibration of hybrid
air-shower detectors, which combine a surface detector array and a fluorescence detector, to obtain an unbiased estimate of the calibration curve.
The special features of calibration data from air showers prevent unbiased results, if a standard least-squares fit is applied to the problem. We develop a general maximum-likelihood approach, based on the detailed statistical model, to solve the problem.
Our approach was developed for the Pierre Auger Observatory, but the applied
principles are general and can be transfered to other air-shower experiments,
even to the cross-calibration of other observables.
Since our general likelihood function is expensive to compute, we derive two
approximations with significantly smaller computational cost. In the recent
years both have been used to calibrate data of the Pierre Auger Observatory. We
demonstrate that these approximations introduce negligible bias when they are applied to simulated toy experiments, which mimic realistic experimental conditions.
\end{abstract}

\maketitle

\section{Introduction}

\begin{figure}[t]
\includegraphics[width=\columnwidth]{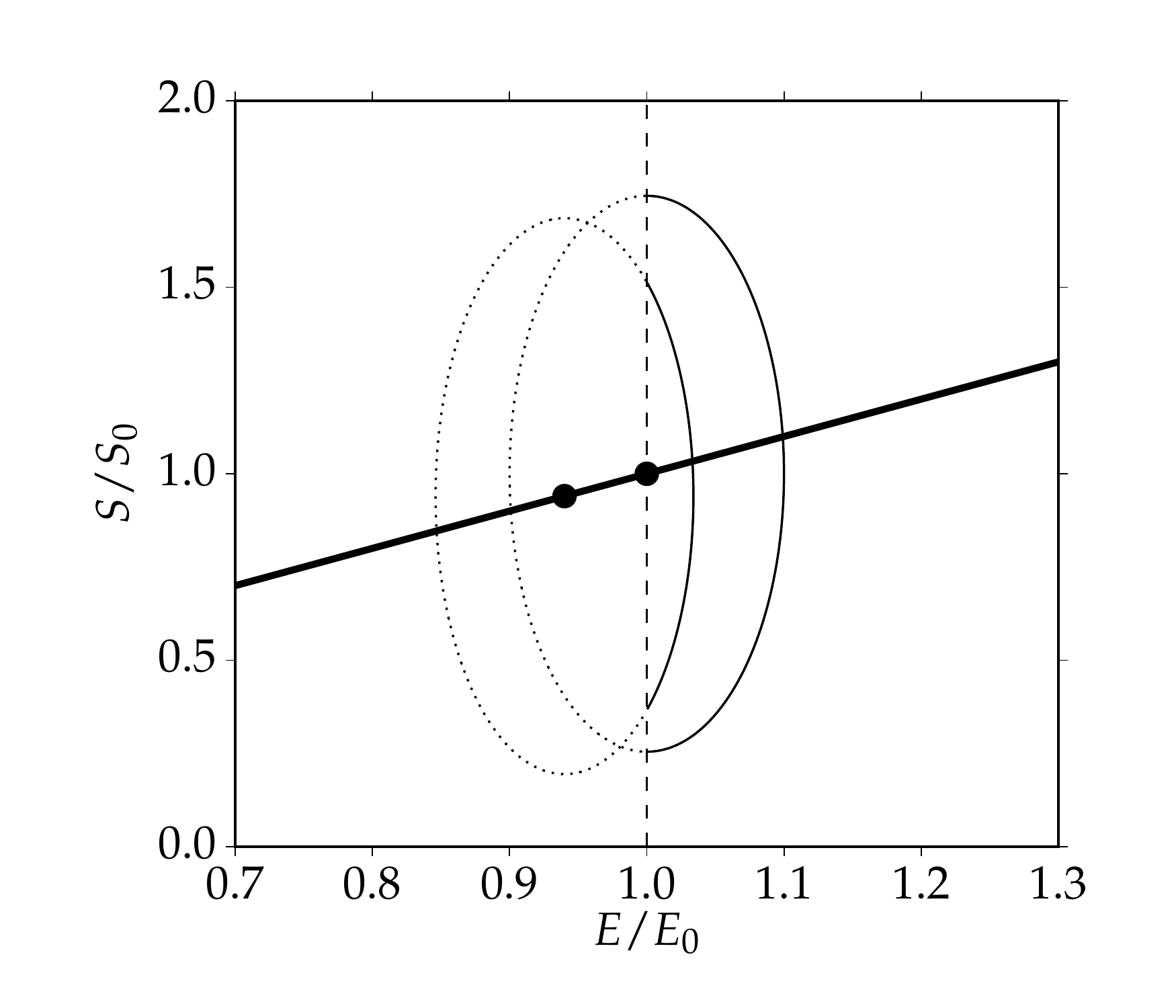}
\capt{fig0}{Sketch of the bias introduced by a simple energy cut.
Shown is an ideal calibration curve $S \propto E$ of the size $S$ of an air shower against its energy $E$ (thick solid line), together with a low energy cut at $E_0$.
Measured estimates of $E$ and $S$ fluctuate around the true values and spread
events from an ideal point on the calibration curve (black dots) outwards into
an uncertainty ellipse.  The line density of events on the ellipse is constant.
Events that migrate below the cut are discarded (dotted arc).  Surviving events
(thin solid arc) spread more often below the ideal calibration curve since the
arc is longer. A least-squares fit wrongly compensates this by placing the
calibration curve below the true curve.}
\end{figure}

The latest generation of air-shower detectors, the Pierre Auger
Observatory~\cite{Auger2008,Auger2010a} and Telescope Array~\cite{Kawai2008},
are hybrid instruments. They combine a fluorescence detector which measures the
calorimetric energy of an air shower with a low duty cycle, and a surface detector array
measuring its size at ground with a full duty cycle.

The size of an air shower measured at the same point in its longitudinal
development is proportional to a power of its energy~\cite{Matthews2005}. Therefore, a calibration function returning an
energy estimate for a measured size can be found by analyzing a subset
of coincident events recorded in both detectors.

Fitting the calibration function to pairs of energy and size estimates
with a plain least-squares method yields biased results for several reasons. Firstly,
the least-squares approach requires the true energy of the air shower to be known
event-by-event, but the fluorescence detector only provides an energy estimate
that fluctuates around the true energy. Secondly, the energy spectrum of
cosmic rays is very steep so that most of the data is located near the lower
energy threshold of the detector.

In the threshold region, the detector triggers are not fully efficient. Upward fluctuations have a higher chance of passing the trigger and entering the data set than downward fluctuations. This creates an acceptance bias, so that the mean size of the accepted events does not reflect the true mean size of the original sample.

Applying an energy cut with a minimum energy high enough to avoid the threshold region altogether solves this problem, but it creates a new bias, caused by event migration over the new threshold introduced by the cut. How the bias appears is illustrated in \fg{fig0}. A superficial solution is to use a slanted cut, but determining the angle under realistic conditions, where the resolutions vary with energy and size of the air shower, requires Monte-Carlo simulation of the data~\cite{Maris2008}.

We will show that a probabilistic approach solves the problem in a consistent way. Based on the known properties of air-shower development
and the detectors, we construct a probability density model for the experimental
data. Maximizing the likelihood of the data under this model then yields an
asymptotically unbiased estimate of the calibration curve.

\section{Definition of variables}

We use the variable $S$ for the size of the air shower at the ground, where it is observed by surface detector arrays. The size $S$ depends on the energy $E$ of the air shower,
mass $A$, and geometry $\va$. We use air-shower geometry as a summary term for the orientation and impact point of the air-shower axis. The size $S$ is often obtained by fitting an empirical lateral distribution function to the ground signals~\cite{Auger2010,TA2012}, but other proxies work as well, such as the inferred total number of muons at ground in very inclined showers~\cite{Auger2014}.

Air showers with the same geometry $\va$ and energy $E$ show a fluctuating
size $S$ at the ground. These fluctuations~\cite{Dembinski2009,Hansen2010,Ulrich2011}
are caused by random outcomes of the first few interactions of the air-shower development
and possibly from sampling a random mass $A$ from the mass-distribution of cosmic rays. The mass $A$ is usually not exactly known event-by-event and therefore the dependency $S(A)$ adds to the observed fluctuations of $S$. We call these fluctuations combined \emph{intrinsic fluctuations}.

Our aim is to find the function that yields the mean size $\ms$ of the air shower,
averaged over intrinsic fluctuations, as a function of its energy $E$ and
geometry $\va$. The energy dependence is usually modeled well by a power
law $p_0 E^{p_1}$. Our approach does not depend on the exact relationship and therefore we will just refer to $\vp$ as the parameter vector of the function $\ms(E, \va, \vp)$.

We mention the dependence of $\ms$ on the full air-shower geometry $\va$ to treat
the most general case. In practice, the dependency on $\va$ is usually corrected before applying the energy calibration. The correction is either based on air-shower simulations~\cite{TA2012,Auger2014}, or inferred from data, by demanding that the flux of cosmic rays looks isotropic in the corrected size~\cite{Hersil1961}.

The inverse of $\ms(E)$ serves as the energy calibration
function, which provides an energy estimate $E_S$ based on a size $S$ of the air shower. Care must be taken, however, since the random fluctuations of the observed size propagate into the energy estimate. Analyses based on $E_S$ need to take into account, that $E_S$ randomly fluctuates around the true energy $E$ event-by-event, combined with the fact that true energies follow a very steeply falling distribution. This makes it more likely that a particular observed value of $E_S$ was generated by an upward fluctuation of an air-shower of lower energy, than by one with the same or higher energy. If the distribution of energies $E$ is to be measured based on $E_S$~\cite{Auger2010}, unfolding methods can be used~\cite{Cowan1998,Dembinski2013}.

In addition to the effects discussed before, detectors do not measure the energy $E$, size $S$, and geometry $\va$ of the air shower directly. They provide estimates $\eest$, $\sest$, and $\vaest$, that randomly
fluctuate around the true values. These fluctuations are caused by statistical
sampling of air-shower particles in the detector and by variations
in the detector response. An experiment therefore provides a sample of tuples $(\eest_i,
\sest_i, \vaest_i)$ as input for the analysis. We assume that an energy cut
$\eest > \ecut$ is applied to this set which discards events with poor
resolution in the threshold region of the detector.

To distinguish between functions and probability density functions (\pdfs) in
this article, we use the semi-colon in \pdfs to separate the random variables
from the dependent variables. For example, $f(x; p)$ is the probability density
function $f$ of the random variable $x$, whose location and shape depends on $p$.
When integrals over random variables appear, we will not explicitly indicate the limits, except if the integral does \emph{not} cover the physical domain of the variable, for example, $[0, \infty)$ for $E$ and $S$.

We will refer to the normal distributions frequently, and therefore use the
notation $\normal(x; \mu, \sigma)$ to indicate the density
\begin{equation}
\normal(x; \mu, \sigma) =
  \frac{1}{\sqrt{2\pi}\sigma}
  \exp\left(-\frac12 \left(\frac{x - \mu}{\sigma}\right)^2\right).
\end{equation}
In a fully rigorous treatment, we would have to use the truncated normal distribution in most places, where the domain of the variable $x$ is not the full real line. We generally assume that the experimental conditions are such that $\mu / \sigma \gg 0$, so that both distributions approach each other.

\section{Likelihood estimation of the calibration function}

Our fitting method is based on the maximum-likelihood method~\cite{James2006}.
For un-binned continuous data, it states that an estimate of the parameter vector $\vp$ can be found by maximizing the joint \pdf $\mathcal{L}$ of the data under the model considered.
We make a usual substitution and maximize $\lnL$ instead of $\mathcal{L}$,
\begin{equation}\label{eq:lnL}
\lnL(\vp) = \sum_i \ln f(\eest_i,\sest_i,\vaest_i;\vp),
\end{equation}
which is equivalent but easier to handle. The density $f(\eest, \sest, \vaest; \vp)$ models the data distribution as a function of $\vp$. We maximize this sum with standard numerical algorithms to get an estimate $\vpest$ of $\vp$.

If the data density was very high, working with a histogram of the data would be more effective and the log-likelihood would take a different form. Both approaches
can also be combined, so that the former is used in high density regions to speed up the
computation of the sum, an example of such a technique is given in
Ref.~\cite{Dembinski2013}.

The maximum-likelihood approach has a useful property that we will exploit repeatedly. Finding the maximum of $\lnL$ to get the estimate $\vpest$ only involves the first derivative $\nabla_\vp \lnL$. Similarly, computing the uncertainty estimate of $\vpest$ only involves the second derivative. Therefore, any constant factors $c_i$ with $\nabla_\vp c_i = 0$, that appear in the evaluation of $f_i(\vp) = f(\eest_i,\sest_i,\vaest_i;\vp)$, can be dropped without changing these results,
\begin{align}
\lnL(\vp)
  &= \sum_i \ln f_i(\vp)
   = \sum_i \ln c_i \, f'_i(\vp) \nonumber \\
  &= \sum_i \ln c_i + \sum_i \ln f'_i(\vp) \equiv \sum_i \ln f'_i(\vp).
\end{align}
We will use this to avoid the explicit computation of such factors wherever possible.

We now focus on the construction of $f(\eest,\sest,\vaest;\vp)$. The
size function $\ms(E,\va,\vp)$ of the air-shower is at the heart of this \pdf, the crucial point is
to model the random fluctuations of events around this mean.

\subsection{Statistical model of the detection process}

The \pdf $f(\eest, \sest, \vaest; \vp)$ of the observed ensemble is constructed by folding several
conditional \pdfs that model the individual sources of fluctuations with the
\pdf $h(E, \va)$ of the arrival frequencies of air showers at the combined aperture
of the detectors. The \pdf $h(E, \va)$ itself is obtained by normalizing the
product of the cosmic-ray flux $J(E) = \dd N / (\dd E \, \dd A \, \dd t \, \dd
\Omega)$ with the effective aperture $A_\text{eff}(E, A, \va)$ of the combined detectors.
The effective aperture can dependent on energy and mass of the cosmic ray. For example,
a fluorescence telescope can see high-energy air showers from a greater distance, since they
are brighter. We emphasize that for the energy calibration to work, any mass
dependency in the effective aperture of the calibration data needs to be the
same as in the final data set to be calibrated.

We now introduce the fluctuations step-by-step. If the detectors were perfect
and there were no intrinsic fluctuations, we would describe the data with
distribution
\begin{align}\label{eq:f1}
f_1(E,S,\va;\vp) = \delta\big(S - \ms(E,\va,\vp)\big) \, h(E,\va),
\end{align}
where the Dirac $\delta$-distribution states that the observed values follow
the function $\ms(E, \va, \vp)$ exactly. Yet, certain pairs of $\ms$ and $E$
occur more frequently than others, due to the different arrival frequencies
of air showers, quantified by $h(E,\va)$. For a fixed geometry $\va$, \eq{f1} represents a line density embedded into the $(E, S)$ plane. It traces the function $\ms(E, \va, \vp)$
that we want to extract.

By modeling how events are randomly scattered away from the line density,
we develop the connection between the function $\ms(E, \va, \vp)$ and the observed ensemble.
Intrinsic fluctuations are incorporated by replacing the $\delta$-distribution
with a conditional \pdf $s$, obtaining
\begin{align}
f_2(E,S,\va;\vp) = s\big(S; \ms(E,\va,\vp),E,\va\big) \, h(E,\va).
\end{align}
The shape of the \pdf $s$ itself can depend on the air-shower energy $E$ and the geometry $\va$.

Now we add fluctuations caused by the detectors, and regard event loss from online triggers
and a minimum energy cut. These effects are modeled by another conditional probability density
function, the detector kernel $g(\eest, \sest, \vaest; E, S, \va)$. We fold $f_2$ with the
kernel and multiply the result with a Heaviside function $\Theta(\eest -
\ecut)$ to model the effect of the applied energy cut. This yields our first main result
\begin{align}\label{eq:fpdf}
f_3(\eest,\sest,\vaest;\vp) &=
  \Theta(\eest - \ecut) \int \dd E \int \dd S \int \dd \va
\nonumber
\\
  & \quad g(\eest,\sest,\vaest;E,S,\va) \, f_2(E,S,\va;\vp).
\end{align}
We neglect here, that shower-to-shower fluctuations in the size $S$ may be accompanied by anti-correlated fluctuations in the energy estimate $\eest$. The energy estimate $\eest$ is typically based on light generated primarily by the electromagnetic cascade. A fraction of the shower-to-shower fluctuations is caused by variations in the flow of cosmic-ray energy into the hadronic and electromagnetic cascade. The anti-correlations reflect the conservation of energy. Anti-correlated fluctuations in the energy estimate are expected to be smaller than 4\,\% above $10^{18}$\,eV, and decreasing at higher energies~\cite{Tueros2013}. This is typically small compared to the energy resolution, and therefore not included in the model.

Apart from this simplification, \eq{fpdf} is a general statistical model of the calibration data. The detector kernel $g(\eest,\sest,\vaest;E,S,\va)$ can be obtained from
Monte-Carlo simulation or derived from an empirical model. The \pdf
$s$ needs to be estimated from air-shower simulations or fitted to the
calibration data together with the function $\ms(E, \va, \vp)$.

Due to losses modeled by the Heaviside function and the detector kernel, $f_3$
is not normalized to unity. The maximum-likelihood method does not require
$f_3$ to be normalized, if the normalization does not depend on the parameters
$\vp$ that are optimized. Otherwise, $f_3$ needs to be replaced by
\begin{equation}\label{eq:fnorm}
f(\eest,\sest,\vaest;\vp) =
  \frac{f_3(\eest,\sest,\vaest;\vp)}
  {\int\dd\eest\int\dd\sest\int\dd\vaest \, f_3(\eest,\sest,\vaest;\vp)}.
\end{equation}
A numerical computation of the normalization is expensive and should be
avoided. We will show how the computation can be neglected in good approximation if a sufficiently large value of $\ecut$ is chosen in the next section.

Our approach in its general form is more complex than a plain least-squares
fit of the data, but if it is a complete probabilistic model of the data,
maximizing the likelihood of the data under the model is guaranteed to yield
an asymptotically unbiased estimate of the calibration curve. In particular,
the probabilistic model handles inefficiencies and event migration above the
threshold defined by the energy cut $\ecut$, which cannot be dealt with in the
framework of least-squares fitting.

The general form of \eq{fpdf} is expensive to compute due to the many integrals.
We will proceed to discuss valid approximations which greatly reduce the
computational cost, up to a point of removing all integrals. We will illustrate
these approximations along an fully fledged example: the application of the
likelihood approach to the energy calibration in the Pierre Auger Observatory.

\section{Application to the Pierre Auger Observatory}

Two variants of our approach have been used in recent analyses from the Pierre
Auger Observatory. The variants are obtained by approximating \eq{fnorm} in a
controlled manner, which reduces it into a practical form.

Both variants have a few aspects in common. The Auger surface detector array is
sufficiently flat and regular, so that the dependence on the air-shower geometry
$\va$ reduces only to a dependence on the zenith angle $\theta$. The remaining
atmospheric attenuation is corrected either empirically by demanding flux isotropy~\cite{Hersil1961, Pesce2011}
or based on air-shower simulations~\cite{Dembinski2010}. Thus, in this case the refined size
parameter $S$ depends only on the energy $E$ of the air-shower, and we have
\begin{equation}
\ms(E) = p_0 E^{p_1}.
\end{equation}

\subsection{Detector kernel}

The detector kernel is factorized into two independent normal distributions for
$\eest$ and $\sest$, and a $\delta$-distribution for $\thest$,
\begin{equation}
g(\eest,\sest,\thest;E,S,\theta) \approx
  g_\eest(\eest;E) \, g_\sest(\sest;S,\theta) \, \delta(\thest - \theta),
\end{equation}
with
\begin{align}
g_\eest(\eest;E) &= \normal\big(\eest; E, \sigma_\eest(E)\big) \\
g_\sest(\sest;S,\theta) &= \normal\big(\sest; S, \sigma_\sest(S, \theta)\big).
\end{align}
Using independent normal distributions for $\eest$ and $\sest$ is well
motivated, since the measurements are practically independent. The
measurements implicitly sum up many individual signals with near-normal distributions.
Due to the central-limit theorem, the resolutions turn Gaussian.

Using the $\delta$-distribution for $\thest$ is an approximation that saves a numerical integration.
The $\delta$-distribution is the limit of a normal distribution with vanishing width, so effectively this ansatz treats the measurement of $\theta$ as exact. For the Pierre Auger Observatory this approximation is very good, since the angular resolution is high and the detector kernel is only a slowly varying function of $\theta$.

Event losses in the detector are neglected. This can be made into a good approximation by choosing a sufficiently high value of the energy cut $\ecut$. The value of $\ecut$ needs to be high enough so that all accepted events have 100\,\% detection efficiency, and rejected events with a reasonable chance to migrate over the threshold still have efficiencies near 100\,\%.

Under these conditions, the normalization of \eq{fpdf} becomes independent of the choice of $\vp$. Therefore, \eq{fpdf} can be used directly in the likelihood instead of \eq{fnorm}. We will drop the Heaviside function $\Theta(\eest - \ecut)$ from now on, since it is always one for the selected data and was only needed to compute the normalization in the general case.

\subsection{Intrinsic fluctuations}

The \pdf $s(S; \ms, E)$ of the intrinsic fluctuations is modeled by a
normal distribution
\begin{equation}
s(S; \ms(E, \vp), E, \vp) = \normal\big(S; \ms(E, \vp), \sigma_S(E)\big).
\end{equation}
In case of the Pierre Auger Observatory, there is no indication that the shower-to-shower fluctuations depend on the air-shower geometry $\va$, so the dependency is dropped.

The choice of a normal distribution is only empirically motivated, since the \pdf is theoretically
unknown. Its shape depends on the unknown distribution of cosmic-ray masses.
However, pure samples of simulated proton and iron air-showers show distributions
close to normal~\cite{Dembinski2009}, and the model fits data from
the Pierre Auger Observatory well~\cite{Pesce2011,Auger2015}.


\subsection{Combined fluctuation model}\label{sec:comb_fluc}

Since both the detector-generated fluctuations $(\sest - S)$ and the intrinsic
fluctuations $(S - \ms)$ are modeled by normal distributions, it is tempting to
carry out the folding
\begin{equation}\label{eq:gs}
g_s(\sest) = \int\dd S\, g_\sest(\sest; S, \theta) \, s(S; \ms(E, \vp), E, \theta)
\end{equation}
analytically, to save another numerical integration. Unfortunately, the trivial solution that holds for two independent normal
distributions
\begin{align}\label{eq:gns}
g_s^\text{N}(\sest) = \normal\big(\sest; \ms(E, \vp), (\sigma^2_\sest + \sigma^2_S)^{1/2} \big)
\end{align}
does not work here, since the resolution $\sigma_\sest$ of the detector depends on the random outcome $S$ from shower-to-shower fluctuations. The fluctuations are coupled.

\begin{figure}
\includegraphics[width=\columnwidth]{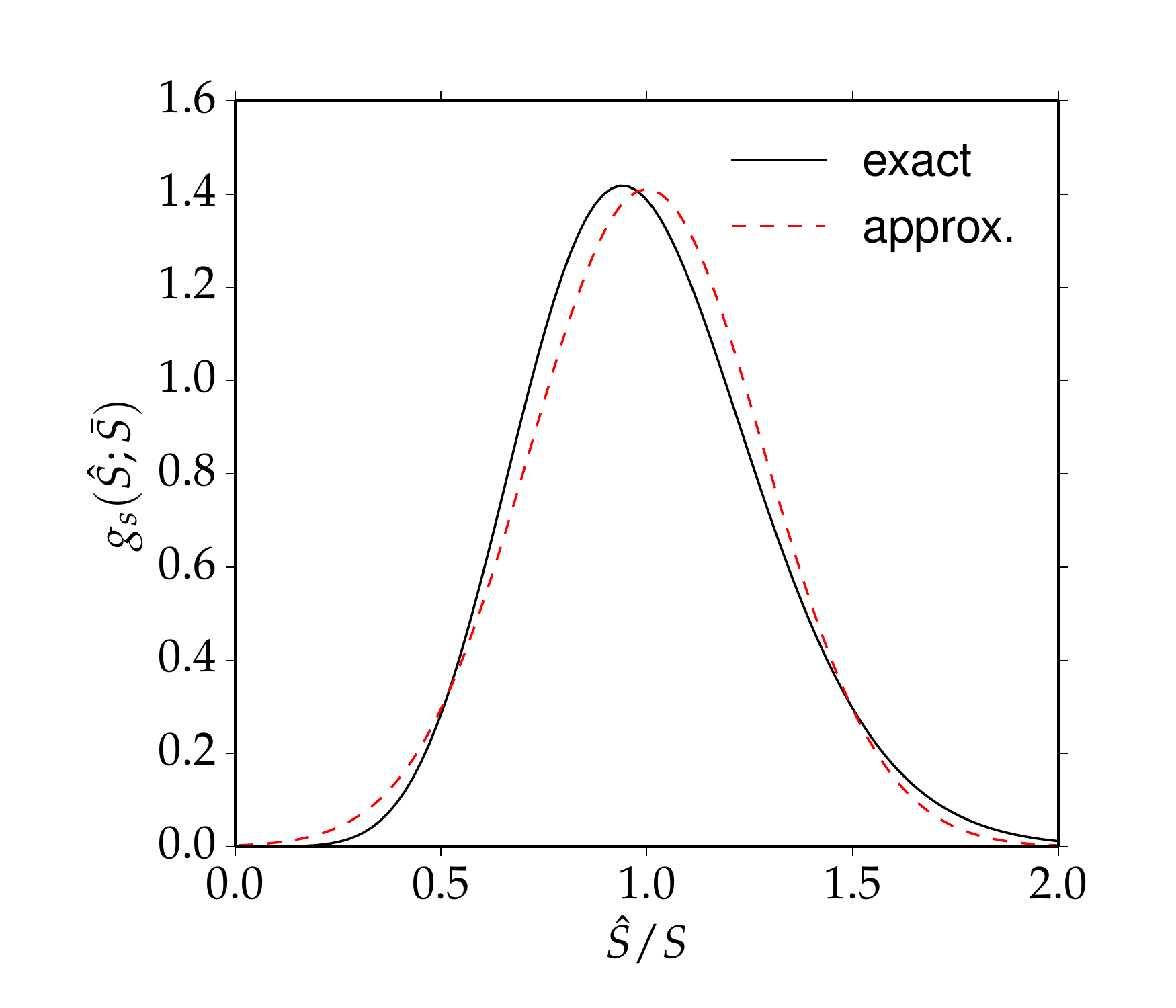}
\capt{fig1}{Exact solution $g_s(\sest; \ms)$ and normal approximation
$g_s^\text{N}(\sest; \ms)$ to the folding of two normal distributions
$g_\sest(\sest - S)$ and $s(S - \ms)$ with $\sigma_\sest / S = \sigma_S / \ms =
0.2$.}
\end{figure}

The difference between the normal approximation $g_s^\text{N}$, obtained by
replacing $S$ with $\ms$ in the computation of $\sigma_\sest$, and the exact
solution $g_s$ is illustrated in \fg{fig1} for realistic values of
$\sigma_\sest / S = \sigma_S / \ms = 0.2$. The exact solution has a shape that
resembles something between a normal and log-normal distribution, in agreement with early studies
based on Monte-Carlo simulations of air showers~\cite{Maris2008}.

The normal approximation $g_s^\text{N}$ is nevertheless acceptable for the
estimation of $\ms(E,\vp)$, since $g_s^\text{N}$ and $g_s$ have the same
expectation value, as pointed out in Appendix~\ref{app:expectation}. The maximum-likelihood estimator of the central value of $g_s^\text{N}$ is unbiased with respect to the expectation value, and therefore this approximation does not bias the fit of the size function $\ms(E,\vp)$.

However, the approximation biases the estimate of the width $\hat{\sigma}_S$ of
the intrinsic fluctuations. While this parameter is irrelevant for the energy
calibration itself, it has to be fitted to data since its true value is unknown.
The width is also of physical interest in its own, since it is
sensitive to the cosmic-ray mass composition. The impact on $\hat{\sigma}_S$ is
shown in Section~\ref{sec:toymc}.

By assembling all pieces together and carrying out the integration over
$\theta$, we arrive at the following form
\begin{multline}\label{eq:variantbase}
f(\eest, \sest, \thest; \vp) =
  \int
    \dd E\, g_\eest(\eest; E) \,
    g_s^\text{N}(\sest; E, \thest, \vp) \, h(E,\thest).
\end{multline}
This form is the common root of two variants which have been independently derived and used for different zenith angle ranges at the Pierre Auger Observatory. Both variants assume that the distribution of hybrid events $h(E, \theta)$ factorizes,
\begin{equation}
h(E, \theta) = h_E(E) \, h_\theta(\theta),
\end{equation}
which follows if the effective aperture $A_\text{eff}$ for hybrid events factorizes in these variables. This was found to hold in good approximation~\cite{Dembinski2009}.

\subsection{Variant A}

In variant A, \eq{variantbase} is integrated over the estimated zenith angle $\thest$, which effectively removes one dimension of the data from the likelihood and the inference. This variant has been used at the Pierre Auger Observatory to calibrate air showers with zenith angles up to $60^\circ$~\cite{Pesce2011}.

The integration leads to an effective kernel on the right hand side of \eq{variantbase}
\begin{equation}
\tilde g_s^\text{N}(\sest; E, \vp) = \int\dd\thest\, g_s^\text{N}(\sest; E, \thest, \vp) \, h_\theta(\thest),
\end{equation}
which is again approximated by a normal distribution. This approximation is good where the zenith angle dependency of the detector kernel is weak.

The remaining partial distribution $h_E(E)$ is approximated by a power law $E^{-\gamma}$ with an appropriate spectral index. We arrive at variant A,
\begin{multline}\label{eq:varianta}
f_A(\eest, \sest; \vp) = C
  \int\dd E\, \frac1{\sigma_E\,\tilde\sigma_\sest} \times \\
    \exp \Bigg(
    -\frac12 \bigg( \frac{\eest -  E}{\sigma_E} \bigg)^{2}
    -\frac12 \bigg( \frac{\sest - p_0 E^{p_1}}{\tilde\sigma_\sest} \bigg)^{2} \Bigg)\,
    E^{-\gamma},
\end{multline}
where $\tilde\sigma_\sest(E) = (\sigma_\sest(p_0 E^{p_1})^2 + \sigma_S(E)^2)^{1/2}$ is an effective resolution that includes shower-to-shower fluctuations, and $C$ is an unspecified normalization constant that does not depend on $\vp$ and is therefore irrelevant for the likelihood estimation.

The log-likelihood function then is (up to constants)
\begin{multline}\label{eq:lnLA}
\lnL_A(\vp) =
  \sum_i \ln \int\dd E\, \frac1{\sigma_E\,\tilde\sigma_\sest} \times \\
    \exp \Bigg(
    -\frac12 \bigg( \frac{\eest_i -  E}{\sigma_E} \bigg)^{2}
    -\frac12 \bigg( \frac{\sest_i - p_0 E^{p_1}}{\tilde\sigma_\sest} \bigg)^{2} \Bigg)\,
    E^{-\gamma}.
\end{multline}
The integration over the true energy $E$ is carried out numerically for each observed tuple $(\eest_i, \sest_i)$.

\subsection{Variant B}

In variant B, the approximations after \eq{variantbase} take a different path. The zenith angle dependency is kept, and a bootstrap estimate is used to remove the last remaining integral. The result is a double sum, which is fast to compute for small to moderate data sets. Variant B has been used to calibrate very inclined events with zenith angles beyond $65^\circ$~\cite{Dembinski2011}.

We start by observing that the \pdf $h_E(E)$ of hybrid events is well
approximated by the \pdf $h_\eest(\eest)$ in the range of interest, which describes the distribution of the observed energy estimates $\eest$. Both \pdfs differ by an integration over the detector kernel $g_\eest$ of the energy
measurement,
\begin{equation}
h_\eest(\eest) = \int \dd E \, g_\eest(\eest; E) \, h_E(E).
\end{equation}
The effect of the folding is small since the resolution of
the energy measurement is about 10\,\%~\cite{Auger2010a}. In particular,
in the region near $\ecut$ and above, the \pdfs differ mainly by a shift that can be
absorbed into the normalization.

After establishing this, we can now estimate the \pdf $h_\eest$ from the calibration data itself. This could be done with a kernel density estimate or another non-parametric density estimation method. Since $h_\eest$ appears only inside an integral, we chose the even simpler bootstrap estimate~\cite{Efron1993}. The bootstrap estimate can be regarded as a kernel density estimate with a kernel of vanishing width. It is constructed as a sum of $\delta$-distributions positioned at the observed values $\eest_k$, weighted by the inverse of the overall detection efficiency $\epsilon_k = \epsilon(\eest_k)$ at this value,
\begin{equation}\label{eq:bootstrap}
h_\eest^\text{B}(\eest) = \frac1K \sum_k \frac{\delta(\eest - \eest_k)}{\epsilon_k}.
\end{equation}
The sum over $k$ is independent of the sum in \eq{lnL} and runs over all $K$ hybrid events, not only those above the cut value $\ecut$.

Bootstrap theory is not well known in physics, but its is an established branch of statistics. For example, the sample mean can be derived as the bootstrap estimate of the expectation value for an arbitrary \pdf $f(x)$
\begin{align}
\expect{x} &= \int\dd x\, x\, f(x)  \rightarrow \text{E}^\text{B}[x] = \int\dd x\, x\, f^\text{B}(x) \nonumber \\
&= \int\dd x\, x\, \frac1K \sum_k \delta(x - x_k) = \frac1K \sum_k x_k.
\end{align}

By inserting \eq{bootstrap} into \eq{variantbase} and integrating over $E$, we
obtain
\begin{multline}\label{eq:variantb}
f_B(\eest, \sest, \thest; \vp) =
  \frac{h_\theta(\thest)}{2\pi K} \times \\
  \sum_k \frac1{\epsilon_k} \normal \big(\eest; \eest_k, \sigma_{\eest, k}\big) \,
         \normal \big(\sest; p_0 \eest_k^{p_1}, \tilde\sigma_{\sest, k} \big),
\end{multline}
with kernel width functions evaluated at the bootstrap values $\eest_k$,
\begin{align}
\sigma_{\eest,k} &= \sigma_\eest(\eest_k) \\
\tilde\sigma_{\sest, k} &= \Big(\sigma_\sest \big( p_0 \eest_k^{p_1}, \thest \big)^2 + \sigma_S(\eest_k)^2 \Big)^{1/2}.
\end{align}

A final approximation replaces the output of the kernel width
functions $\sigma_\eest$ and $\sigma_\sest$ with their event-wise estimates $\hat\sigma_{\eest,k}$ and $\hat\sigma_{\sest,k}$ from the reconstruction algorithms. This change is not strictly necessary, but it avoids the need to parametrize and fit the kernel width functions to data in advance. This leaves the efficiency function $\epsilon(\eest)$ as the only external input.

Inserting \eq{variantb} into \eq{lnL} yields a curious double sum over the
energy estimates, caused by the bootstrap approximation. After dropping all constant factors that do not depend on $\vp$, which includes $h_\theta(\thest)$, we obtain

\begin{multline}\label{eq:lnLB}
\lnL_B(\vp) =
  \sum_i \ln \left[
    \sum_k \frac1{\hat\sigma_{\eest,k} \, \hat{\tilde\sigma}_{\sest,k} \, \epsilon_k}
  \times \right.
\\
\left. \exp\left(
  -\frac12
  \left(
    \frac{\eest_i - \eest_k}{\hat\sigma_{\eest,k}}
  \right)^2 -
  \frac12
  \left(
    \frac{\sest_i - p_0 \eest_k^{p_1}}{\hat{\tilde\sigma}_{\sest,k}}
  \right)^2
\right) \right],
\end{multline}
where the sum over $i$ only includes events with $\eest_i > \ecut$. This is the final result of approximation B.

The structure of \eq{lnLB} allows for some optimization, which may make the numerical maximization of the log-likelihood faster. The matrix
$z_{ik}=(\eest_i-\eest_k)/\sigma_{\eest,k}$ is constant with respect to changes
in $\vp$, and can be precomputed. Terms with $|z_{ik}|>8$ can be discarded altogether, as their contributions to the second sum are negligible.

We note that \eq{lnLB} cannot be further reduced into an equivalent
least-squares method due to the sum inside the logarithmic terms.

\section{Performance in toy simulations}
\label{sec:toymc}

We test the frequentistic properties of our fits on a set of simulated experiments.
We are particularly interested in the bias $\expect{\vpest-\vp}$ of the
maximum-likelihood estimate $\vpest$ and the true coverage of the estimated
confidence region. We approximate the confidence region with an ellipsoid in the
usual fashion, represented by the covariance matrix obtained after inverting the Hesse
matrix of $-\lnL(\vp)$.

\begin{figure}
\includegraphics[width=\columnwidth]{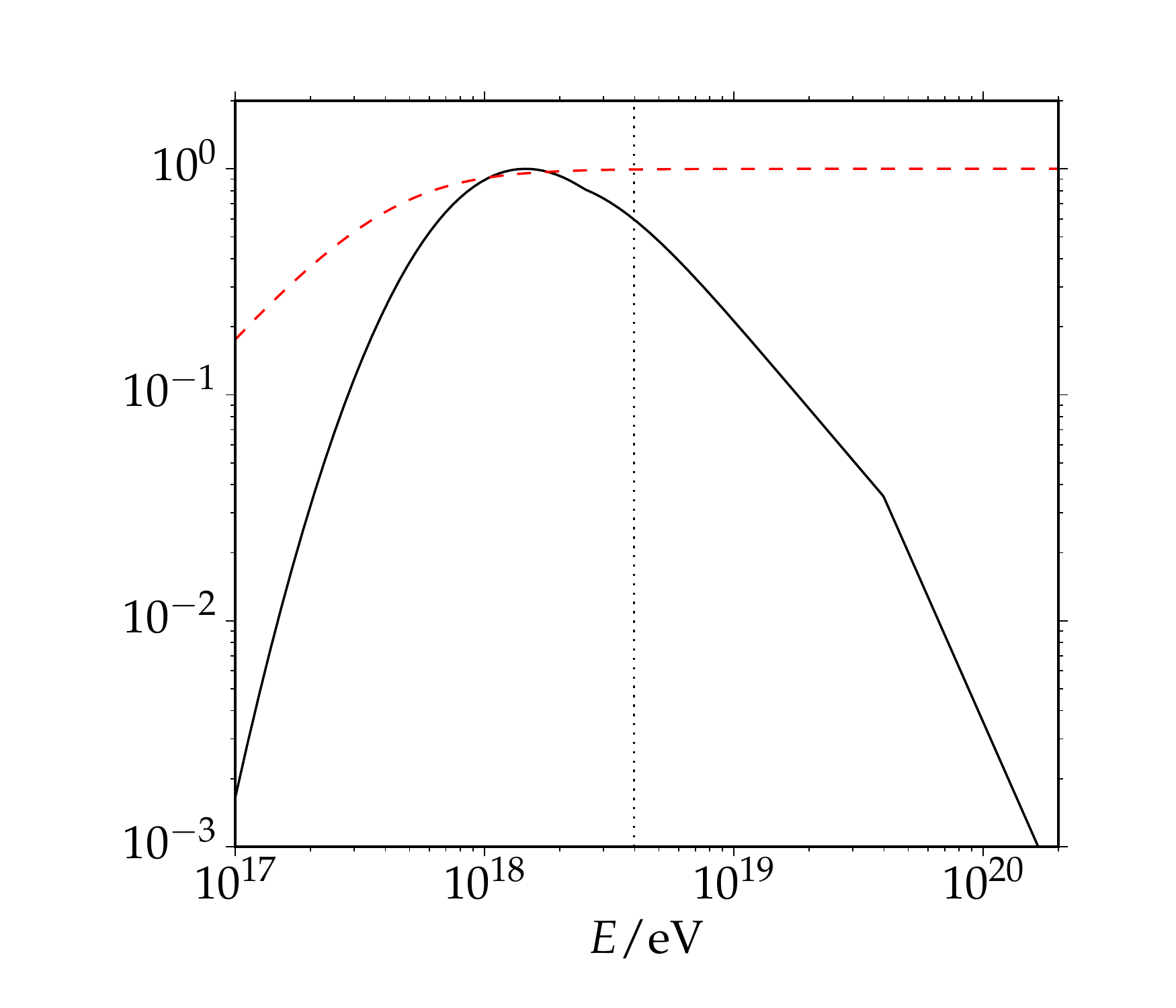}
\capt{fig2}{Energy distribution $\dd N/ \dd \ln E$ of hybrid events (solid line) in arbitrary units and effective trigger efficiency (dashed line) used in the toy simulation. The minimum energy cut used in the fits is also indicated (dotted line).}
\end{figure}

In order to study these statistical properties, we simulate data sets of
toy experiments. In these simple simulations data points are drawn
from parametrized distributions, which mimic the experimental conditions
for the detection of very inclined air showers at the Pierre Auger Observatory.
The results obtained here hold equally well for a simulation of less inclined
events. The parametrization were taken from a previous study~\cite{Dembinski2009}
and are summarized in Appendix~\ref{app:toymc}. The distribution of hybrid events and
the average event loss due to the simulated trigger is shown in \fg{fig2}.

We generate 1000 toy experiments. In each toy experiment, we generate
events until 200 pass the energy cut $\ecut=10^{18.6}$\,eV. The
true size function is taken to be
\begin{equation}\label{eq:cal}
\ms(E) = p_0 \, (E/10^{19}\,\text{eV})^{p_1}
\end{equation}
with input values $\vec{p}=(2.0, 0.9)$.

We fit the toy data with both variant A and B, using \eq{lnLA} and \eq{lnLB}, respectively.
For each simulated experiment and variant, we obtain a parameter vector $\vpest$.
After fitting many independent experiments, the average $\langle \vpest - \vp \rangle$
will approach the bias.

In case of variant A, we insert the true resolution functions used in the toy simulation in the integrand of \eq{lnLA}. The spectral index is set to $\gamma = -2.4$, based on a power law fit to the simulated hybrid distribution between $10^{18.5}$\,eV and $10^{19.5}$\,eV.

In case of variant B, we set the efficiencies $\epsilon_k = 1$ in \eq{lnLB} for simplicity, since the effective trigger efficiency does not depart significantly from one above $10^{18}$\,eV. Only this region is potentially relevant for event migration above the energy cut.

Intrinsic fluctuations in the toy simulation have a constant relative resolution,
$\sigma_S / \ms = 0.15$. To obtain results that generalize well, we do not assume
the same in the likelihood fits. In principle, the relative resolution could vary smoothly with energy,
for example, if the mass composition changes with energy. Therefore, we model the
intrinsic fluctuations with a second-order Bernstein polynomial,
\begin{gather}
\sigma_S / \ms =
  q_0\,(1-z)^2 + q_1\,(1-z)\,z + q_2\,z^2
\label{eq:sigmas}
\\
z = \begin{cases}
  0 & \text{; if}\; E < 10^{18}\,\text{eV} \\
  \frac{\lg(E / 10^{18}\,\text{eV})}
       {\lg(10^{20}\,\text{eV} / 10^{18}\,\text{eV})}
    & \text{; if}\; 10^{18}\,\text{eV} < E \le 10^{20}\,\text{eV} \\
  1 & \text{; if}\; E > 10^{20}\,\text{eV}
\end{cases},
\nonumber
\end{gather}
and fit the three parameters $\vq$ along with the parameters $\vp$ to each data set. The parametrization with a Bernstein polynomial allows us to implement the physical constraint $\sigma_S > 0$ with a simple lower limit $q_\ell > 0$ on the parameters, supported by most numerical optimization algorithms.

While the intrinsic fluctuations are not of primary interest for the calibration curve, they need to be fitted in order to complete the probabilistic model. Assuming that they are zero leads to a significant bias in the fitted parameters $\vp$ of the toy experiments. In addition, fitting the intrinsic fluctuations provides valuable physical information, since they are sensitive to the cosmic-ray mass composition~\cite{Dembinski2009, Ulrich2011}.

To put our likelihood methods in perspective, we also apply a naive least-squares fit,
where $\vpest$ is obtained by minimizing
\begin{equation}
\chi^2(\vp) = \sum_i \left(\frac{\sest_i - p_0 {\eest_i}^{p_1}}{\hat\sigma_{\sest,i}} \right)^2,
\end{equation}
where $\hat\sigma_{\sest,i}$ is the event-wise uncertainty estimate provided by the simulation.

\begin{figure}
\includegraphics[width=\columnwidth]{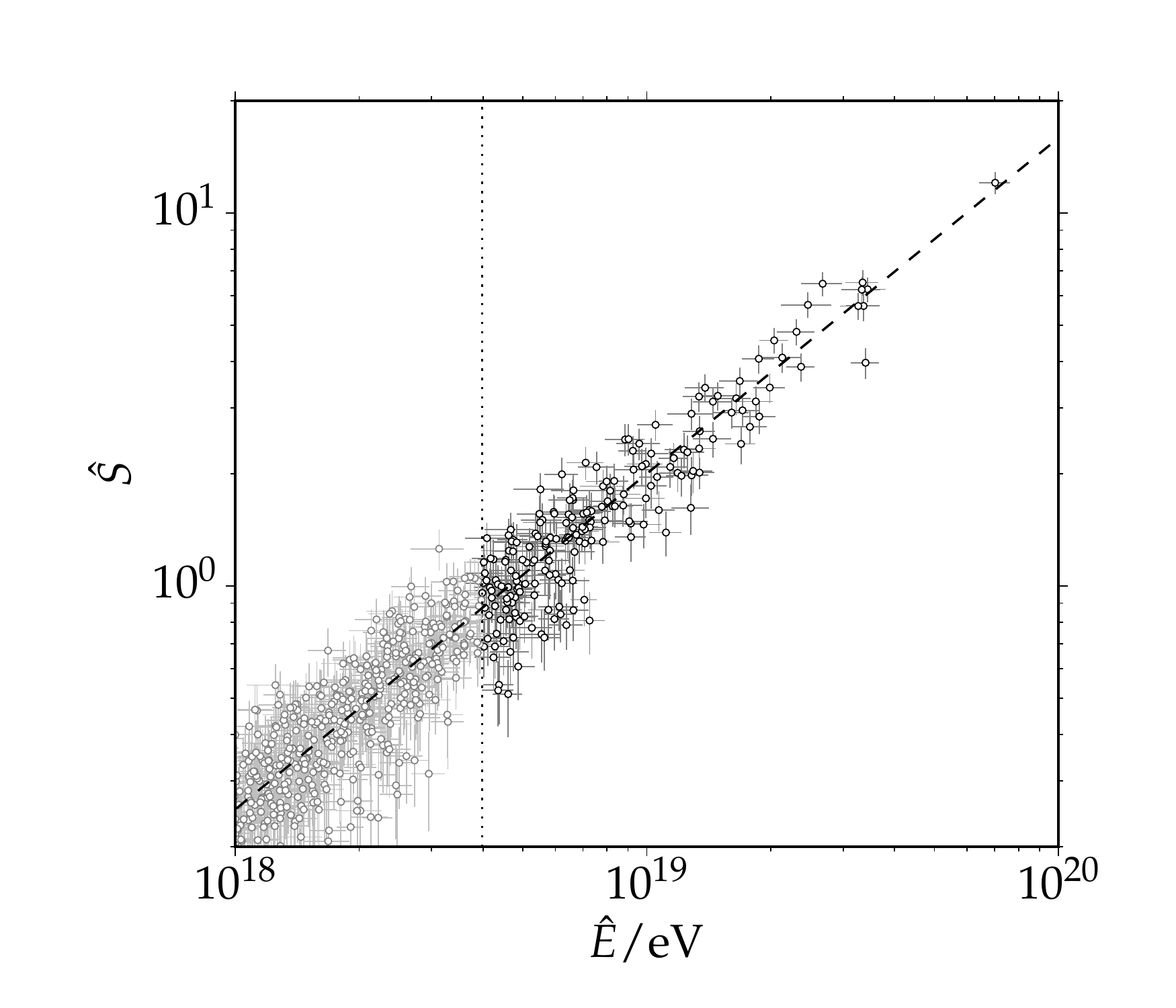}
\capt{fig3}{Simulated data (circles with error bars) from one toy experiment. Also shown are the true mean size $\ms$ (dashed line) of the air showers and the energy cut at $10^{18.6}$\,eV (dotted line).}
\end{figure}

\begin{figure}
\includegraphics[width=\columnwidth]{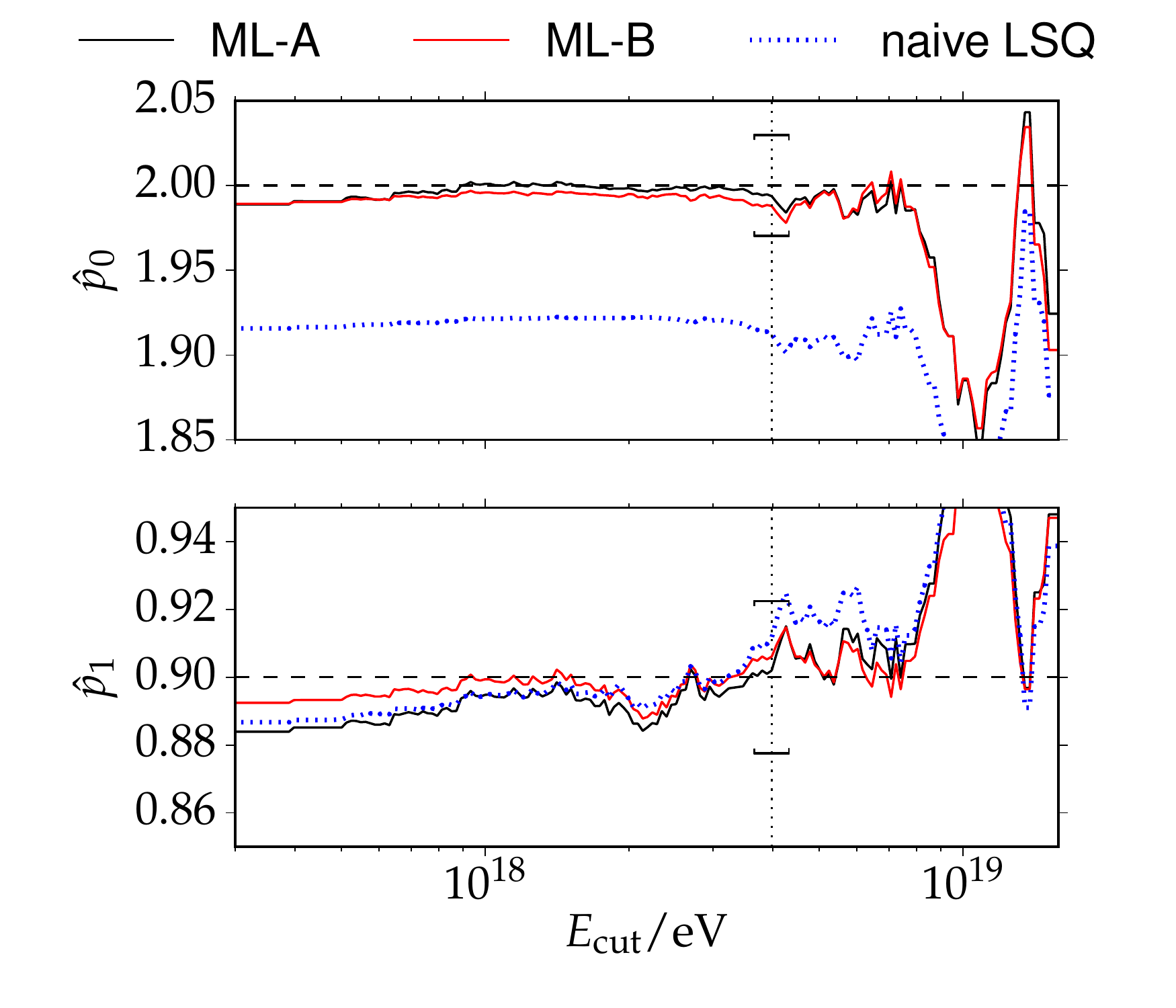}
\capt{fig4}{Fitted parameters for the data shown in \fg{fig3} as a function of the energy cut value, from the likelihood fits (solid lines) and the naive least-squares fit (dotted line). The true parameter values from the simulation (dashed lines) and the nominal cut value are indicated (dotted vertical line). Brackets indicate the statistical uncertainty of the fit at the nominal cut value. The jitter is caused by the discreteness of the data set. It increases with the energy cut, since the number of accepted events decreases and the lever arm for the fit becomes shorter.}
\end{figure}

One of the toy experiments is shown in \fg{fig3}. Only data points above the energy cut value $E_\text{cut}$ enter the fits directly. In case of variant B, the data to the left is still used indirectly to construct the bootstrap estimate of $h_E(E)$, as described in the previous section.

A detailed comparison of the fit results for this data set is shown in \fg{fig4}, which in addition illustrates the dependence on the energy cut. The naive least-squares fit shows a large bias for any cut value in $p_0$, while the likelihood variants appear unbiased for energy cuts higher than the nominal value and give very similar results. A consistent bias for energy cuts below $10^{18}$\,eV is also found for the likelihood fits. The jitter in the scans varies from experiment to experiment, but these biases appear consistently.

The bias in case of low energy cuts is expected, since the trigger efficiency begins to deviates significantly from one below $10^{18}$\,eV, as shown in \fg{fig2}, and the selection bias due to the trigger becomes important. Since efficiency terms were taken out of the statistical model to obtain the approximations, the fits are not applicable at such low energy cuts.

\begin{figure}
\includegraphics[width=\columnwidth]{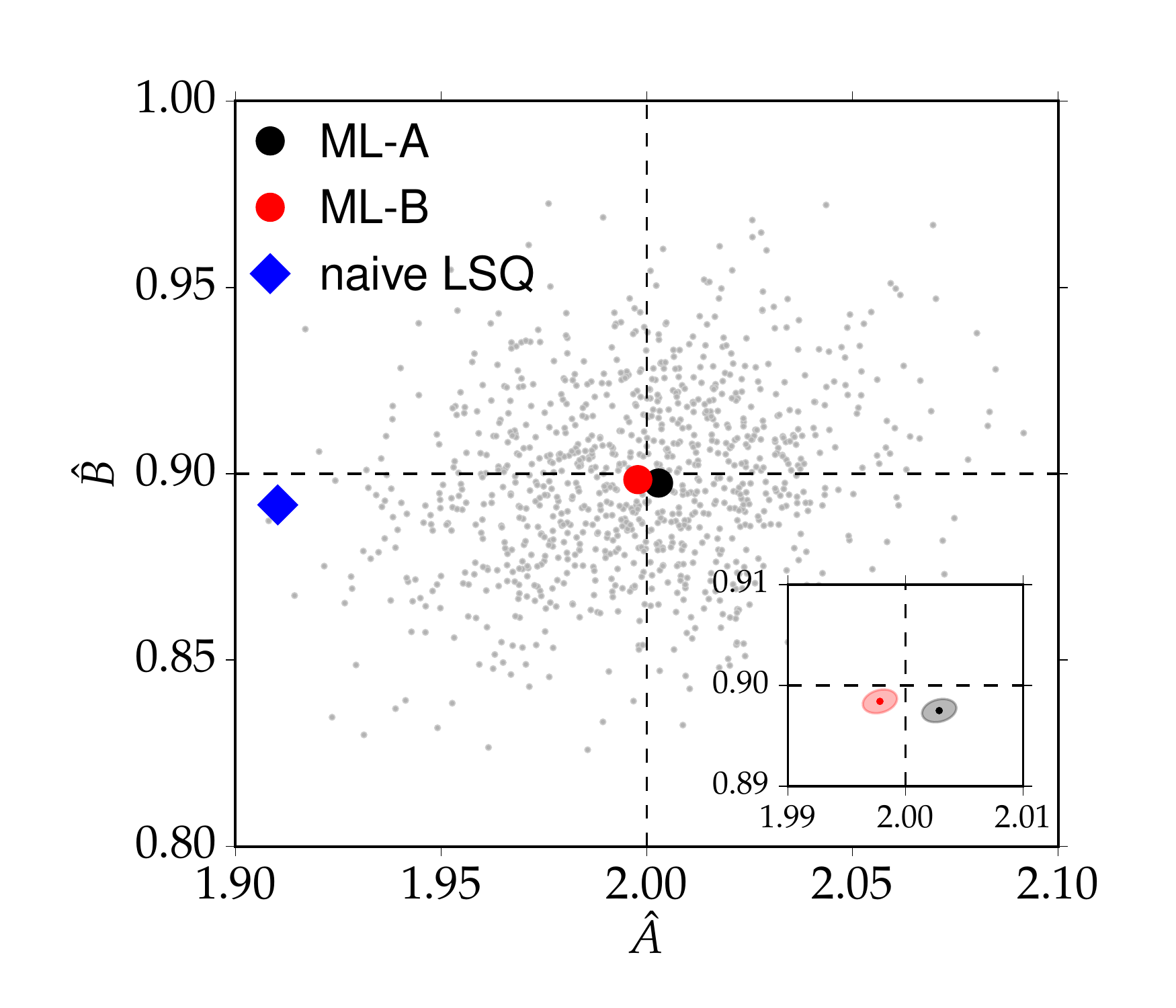}
\capt{fig5}{Averaged parameter estimates from 1000 toy experiments for the two likelihood fits (circles) and the naive least-squares fit (diamond), compared to the true values (dotted lines). Shown in the background are the individual outcomes obtained from variant B (small dots), those of variant A are very similar. The inset zooms closer to the true values. Ellipses around the average fitted values represent the statistical uncertainty of the average.}
\end{figure}

\begin{figure}
\includegraphics[width=\columnwidth]{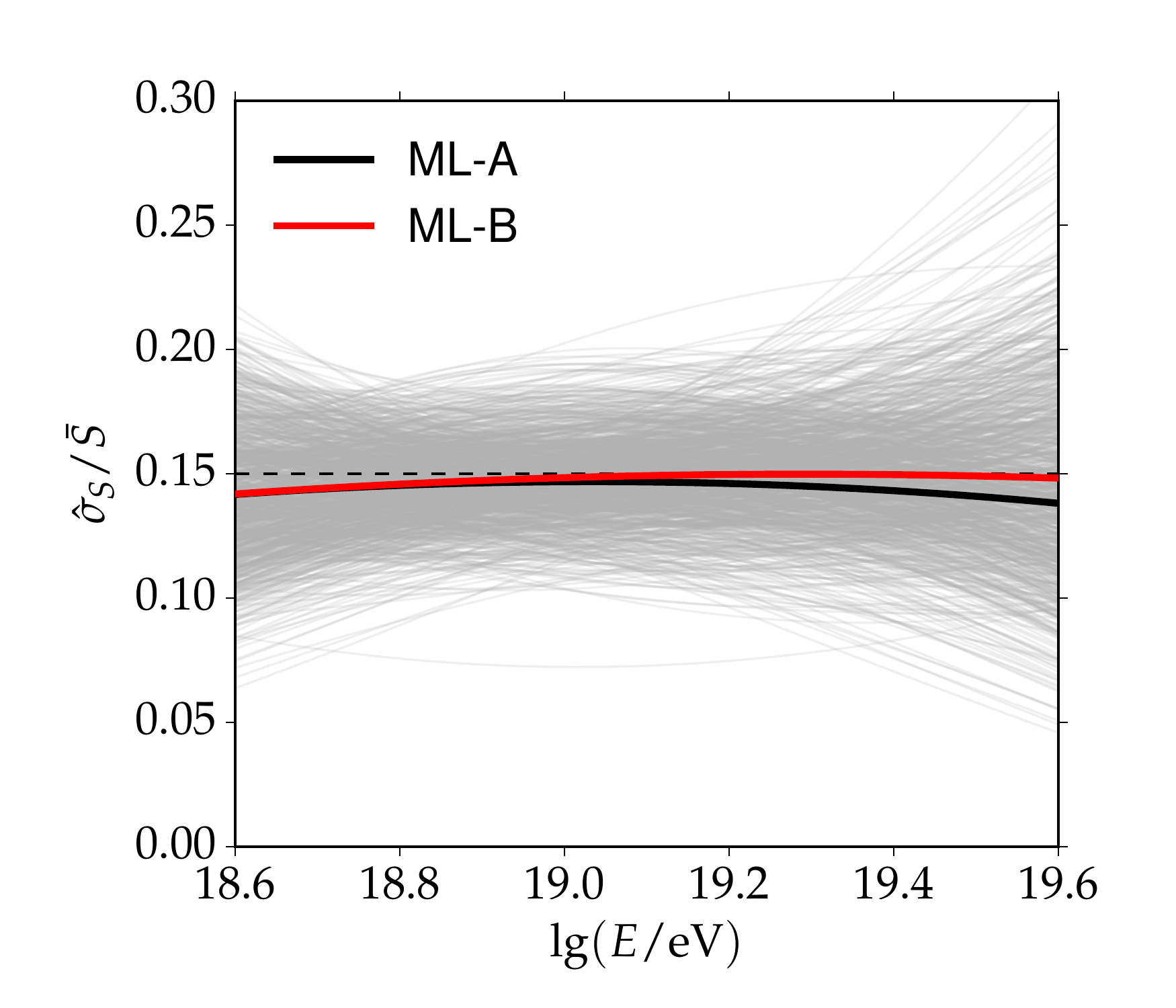}
\capt{fig6}{Averaged estimate of the intrinsic fluctuations $\hat{\sigma}_S / \ms$ from 1000 toy experiments for the two likelihood fits (thick lines with error bands), compared to the true constant value (horizonal dots). Shown in the background are the individual outcomes obtained with variant B (thin lines), those of variant A are very similar.}
\end{figure}

\begin{table}
\begin{tabular}{c c c}
& $\langle \hat p_0 \rangle, \; (\langle \hat p_0 - p_0\rangle)$ & $\langle \hat p_1 \rangle, \; (\langle \hat p_1 - p_1\rangle)$ \\
\hline
ML-A      & $2.003 \pm 0.001 \; (+0.003)$ & $0.898 \pm 0.001 \; (-0.002)$ \\
ML-B      & $1.998 \pm 0.001 \; (-0.002)$ & $0.898 \pm 0.001 \; (-0.002)$ \\
Naive LSQ & $1.910 \pm 0.001 \; (-0.090)$ & $0.892 \pm 0.001 \; (-0.008)$ \\
\end{tabular}
\capt{bias}{Averages of the fitted parameters from 1000 toy experiments for the two likelihood fits and the naive least-squares fit. The estimated bias is shown in parentheses.}
\end{table}

The biases are further explored in \fg{fig5} and \tb{bias}. The naive least-squares fit is strongly biased, while the observed bias in the two likelihood fits is negligible. The statistical uncertainties in a single toy experiment are an order of magnitude larger. The estimated confidence regions constructed from log-likelihood cover the true values in 66\,\% of the toy experiments for variant A and B, very close to the expected 68\,\%.

The fitted intrinsic fluctuations $\hat{\sigma}_S$ are shown in \fg{fig6}. Both likelihood fits show a small bias $\langle \hat\sigma_S/\ms - \sigma_S/\ms \rangle$ of about 10\,\% below $10^{19}$\,eV. At higher energies, variant B performs slightly better than variant A. The variation in the fit from experiment to experiment is large, however, especially at the high end of the energy range. The scatter at the high end reflects that the fit is less constrained where the data density is low. The interpretation of trends in the fitted fluctuations therefore has to be done carefully.

Further exploration showed that the bias vanishes, if the toy simulation is done
with the normal approximation $g^\text{N}_\sest$ to the combined fluctuation
model instead of the correct computation. The bias is therefore a consequence
of the normal approximation described in Section \ref{sec:comb_fluc}.

In a final test, we push the bootstrap estimate of $h_\eest(\eest)$ in variant B to the extreme by generating another set of 1000 toy experiments with only 10 events above $\ecut$. Again, we find only negligible bias in the parameters of the calibration curve.

\section{Conclusion}

We presented a statistical model that describes data taken by a hybrid air-shower detector, consisting of a surface detector array, which measures a size of an air shower, and a fluorescence detector, which measures an energy estimator. We developed a maximum-likelihood approach based on the statistical model, which allows us to infer an asymptotically unbiased energy calibration curve for the size from coincident events observed in both detector parts.

Since the general model is somewhat cumbersome to use, we derived two approximations. The approximations lead to handy formulas, without sacrificing accuracy or introducing significant bias. Both approximations are used by the Pierre Auger Observatory in different zenith angle ranges.

We applied the more aggressive approximation to simulated toy experiments, to investigate the statistical performance. The results showed that the estimated calibration curves are not biased with respect to the true curve used in the generation of the toy data.

\begin{acknowledgments}
We gratefully acknowledge the very fruitful exchanges we had with all of our colleagues in the Auger collaboration. The valuable discussions with Michael Unger need to be highlighted especially. This work was supported in part by the Helmholtz Alliance for Astroparticle Physics HAP, funded by the Initiative and Networking Fund of the Helmholtz Association, by the German Federal Ministry for Education and Research (BMBF), Grant 05A11VK1, and by the ANR-2010-COSI-002 grant of the French National Research Agency. I.C.~Mari\c{s} acknowledges the financial support by the European Community 7th Framework Program, through the Marie Curie Grant FP7-PEOPLE-2012-IEF, no. 328826.
\end{acknowledgments}

\appendix

\section{Expectation of $g_s^\text{N}(\sest;\ms)$ and $g_s(\sest;\ms)$}
\label{app:expectation}

We compute the expectation value of \eq{gs} and \eq{gns}. The experimental conditions that we consider are $\sest / \sigma[\sest] \gg 0$ and $\ms / \sigma[\ms] \gg 0$. This allows us to compute the expectation value over the whole domain of real numbers in good approximation. The expectation value for \eq{gns} immediately follows
\begin{equation}
\expect{\sest} = \int_{-\infty}^\infty \dd \sest \, \sest \, g_s^\text{N}(\sest; \ms) = \ms,
\end{equation}
since $g_s^\text{N}(\sest; \ms)$ is a normal distribution around $\ms$.

For the computation of the expectation value of \eq{gs}, we use that by definition the expectation values of $g_\sest(\sest; S)$ and $s(S; \ms)$ are $S$ and $\ms$, respectively. By changing the order of integration, we obtain
\begin{align}
\expect{\sest} &=
  \int_{-\infty}^\infty \dd \sest\, \sest\,
  \int_{-\infty}^\infty \dd S\, g_\sest(\sest; S) \, s(S; \ms)
\\
  &= \int_{-\infty}^\infty \dd S\, s(S; \ms)
     \int_{-\infty}^\infty \dd \sest\, \sest \, g_\sest(\sest; S)
\\
  &= \int_{-\infty}^\infty \dd S\, s(S; \ms) \, S = \ms.
\end{align}

\section{Toy simulation of calibration data}\label{app:toymc}

We make a Monte-Carlo simulation of \eq{fpdf} to obtain artificial calibration
data. We start by simulating the arrival distribution $h(E, \theta) = h_E(E) \,
h_\theta(\theta)$ of air showers at the detector aperture. The events generated
here simulate highly-inclined events with $60^\circ < \theta < 80^\circ$, as
they are seen by the Pierre Auger Observatory. In the numerical formulas that
follow, the energy $E$ is in units of eV, the zenith angle $\theta$ in units of
radian.

The energy spectrum $h_E$ is modeled by a broken power law with a low-energy
suppression that takes the form of the cumulative normal distribution. The
zenith-angle spectrum $h_\theta$ is modeled as modified exponential decay,
\begin{align*}
h_E(E) &\propto
  \erfc\left(-\frac{\lg E - p_0}{\sqrt{2} p_1}\right)
    \times
\\
  & \quad \times
\begin{cases}
  E^{p_2 - 0.3} & \text{; if } 17.0 < \lg E \le 18.3, \\
  E^{p_2}       & \text{; if } 18.3 < \lg E \le 19.6, \\
  E^{p_2 - 1.2} & \text{; if } 19.6 < \lg E,
\end{cases}
\\
h_\theta(\theta) &\propto
  \exp(p_3\,z + p_4\,z^2), \quad z = \theta - 1.047,
\\
\vp &= (18.3, 0.3, -2.3, -6.4, -45.0).
\end{align*}

The shower-to-shower fluctuations are taken to be normal-distributed and with
a constant relative resolution $\sigma_S / \ms = 0.15$.

The detector-generated fluctuations for the energy estimate $\eest$ and the
size estimate $\sest$ are drawn from normal distributions with relative
resolutions,
\begin{align*}
\sigma_\eest / E &=
\begin{cases}
p_0 + p_1(\lg E - 18.4)^2 & \text{; if } \lg E \le 18.4, \\
p_0 & \text{; otherwise},
\end{cases}
\\
\sigma_\sest / S &= p_2 + p_3/\sqrt{S},
\\
\vp &= (0.10, 0.03, 0.04, 0.10).
\end{align*}
The reconstruction codes in the real experiment provide event-wise estimates
of these resolutions, which are used by fits. The estimates randomly vary
around the true resolutions. To simulate this, we multiply the true resolutions
that are used internally with a factor $(1 + 0.1 z)$
before they are passed on to the fits, where $z$ is standard normal-distributed.

The reduced trigger efficiency for small air-shower sizes is included in the toy
simulation. The trigger probability has the form of a cumulative normal
distribution,
\begin{align*}
P(\sest, \theta) &=
  \frac12
  \erfc\left(
    \frac{\lg\sest - \mu(\theta)}{\sqrt2\sigma(\theta)}
  \right),
\\
\mu(\theta) &= (1 - z) p_0 + z p_1, \quad z = \theta / 0.35,
\\
\sigma(\theta) &= (1 - z) p_2 + z p_3, \quad z = \theta / 0.35,
\\
\vp &= (-0.95, -1.3, 0.2, 0.6).
\end{align*}
The probability is taken to be a function of the size estimate $\sest$,
not $S$, since the trigger decision is highly correlated with the sampling
fluctuations of $\sest$ around the true size $S$. The reduced trigger efficiency
of the energy measurement is effectively included in the energy distribution $h_E(E)$
and not simulated separately.

\end{document}